%% file: main.tex
\documentclass[10pt,journal,compsoc]{IEEEtran}
\ifCLASSOPTIONcompsoc
  \usepackage[nocompress]{cite}
\else
  \usepackage{cite}
\fi
\usepackage{graphicx}
\graphicspath{{./figure/}}
\DeclareGraphicsExtensions{.pdf}

\ifCLASSINFOpdf
\else
\fi
\usepackage{amsmath}

\usepackage{amsfonts}
\usepackage[colorlinks]{hyperref}
\usepackage[utf8]{inputenc}
\usepackage{cleveref}
\crefformat{section}{§#2#1#3}

\newcommand{\PHM}[1]{\vspace{.2em}\noindent\textbf{#1}\hspace{.5em}} % paragraph heading in the middle of a section

\usepackage{multirow}
\usepackage{booktabs}
\usepackage{array}
\usepackage{subcaption}

\usepackage{xcolor}
\definecolor{airforceblue}{rgb}{0.36, 0.54, 0.66}
\hypersetup{
  linkcolor  = blue,
  citecolor  = magenta,
  urlcolor   = airforceblue,
  colorlinks = true,
}

\begin{document}
% \hypersetup{linkcolor=blue, citecolor=magenta}
\title{Towards Efficient Synchronous\\Federated Training: A Survey on\\System Optimization Strategies}

\author{
\IEEEauthorblockN{Zhifeng Jiang,~\IEEEmembership{Student~Member,~IEEE,}
        Wei Wang,~\IEEEmembership{Member,~IEEE,}
        Bo Li,~\IEEEmembership{Fellow,~IEEE,}\\
        Qiang Yang,~\IEEEmembership{Fellow,~ACM}% <-this % stops a space
\IEEEcompsocitemizethanks{\IEEEcompsocthanksitem Z.~Jiang, W.~Wang, B.~Li and Q.~Yang are with the Department
of Computer Science and Engineering, the Hong Kong University of Science and Technology,
Clear Water Bay, Kowloon, Hong Kong.
Q.~Yang is also affiliated with WeBank.\protect\\
E-mail: \{zjiangaj, weiwa, bli, qyang\}@cse.ust.hk
\IEEEcompsocthanksitem This version is peer-reviewed and accepted to appear in IEEE Transactions on Big Data (TBD), 2022, with the Digital Object Identifier being 10.1109/TBDATA.2022.3177222.}
}
}% <-this % stops an unwanted space

\IEEEpubid{2332-7790~\copyright~2022 IEEE. Personal use of this material is permitted.  Permission from IEEE must be obtained for all other uses.}


\IEEEtitleabstractindextext{%
\begin{abstract}
\input{./content/abstract.tex}
\end{abstract}

}

\maketitle

\IEEEdisplaynontitleabstractindextext

% \fi
\IEEEpeerreviewmaketitle

\input{./content/introduction.tex}
\input{./content/background.tex}
\input{./content/prior.tex}
\input{./content/benchmark.tex}
\input{./content/conclusion.tex}

% \section{Conclusion}
% The conclusion goes here.

% or


% \appendices

% \section{}

\ifCLASSOPTIONcompsoc
  \section*{Acknowledgments}
\else
  \section*{Acknowledgment}
\fi

The research was supported in part by RGC RIF grant R6021-20, and RGC GRF grants under the contracts 16209120, 16200221, and 16213120. We also thank the anonymous reviewers for their valuable feedback.


\ifCLASSOPTIONcaptionsoff
  \newpage
\fi



\Urlmuskip=0mu plus 1mu\relax
\bibliographystyle{IEEEtran}
\bibliography{./ref.bib}


% \end{thebibliography}

\begin{IEEEbiography}[{\includegraphics[width=1in,height=1.25in,clip,keepaspectratio]{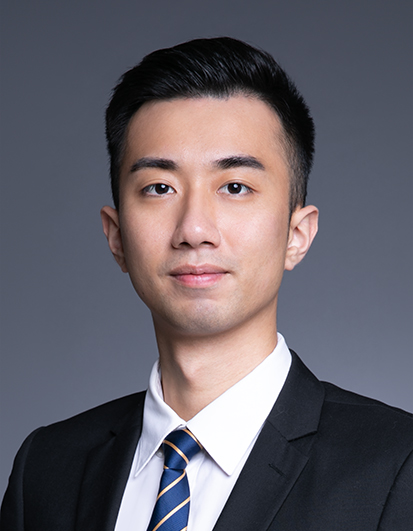}}]{Zhifeng Jiang}
received the BEng (Hons.) degree from the Department of Computer Science and Technology at Zhejiang University (ZJU) in 2019. Since then, he has been working towards a Ph.D. degree in the Department of Computer Science and Engineering, Hong Kong University of Science and Technology (HKUST). He is mainly interested in machine learning systems, with a special focus on efficient and scalable federated learning. He was a recipient of the Best Paper Runner-up Award at IEEE ICDCS 2021. He served on the artifact evaluation committees of ACM SOSP 2021, USENIX OSDI 2022, and USENIX ATC 2022.
\end{IEEEbiography}

\begin{IEEEbiography}[{\includegraphics[width=1in,height=1.25in,clip,keepaspectratio]{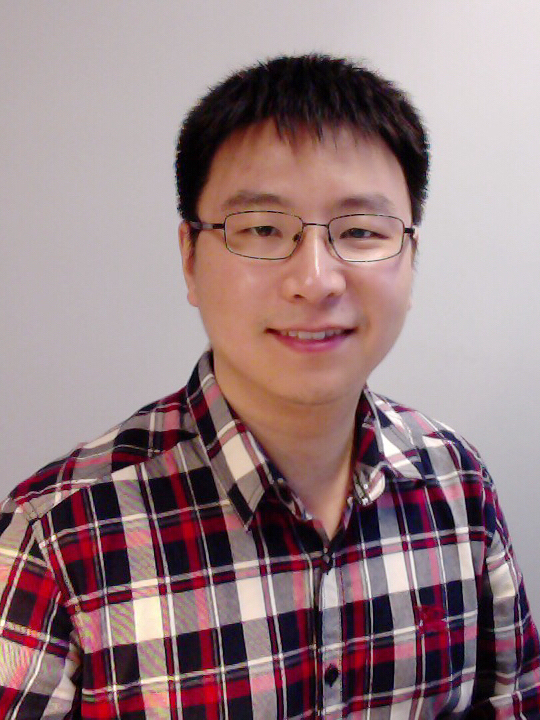}}]{Wei Wang} (Member, IEEE) received his B.Engr. and M.Engr. degrees from the Department of Electrical Engineering, Shanghai Jiao Tong University, China, in 2007 and 2010, respectively and his Ph.D. degree from the Department of Electrical and Computer Engineering, University of Toronto, Canada, in 2015. Since 2015, he has been with the Department of Computer Science and Engineering at the Hong Kong University of Science and Technology (HKUST), where he is currently an Associate Professor. He is also affiliated with the HKUST Big Data Institute. Dr. Wang's research interests cover the broad area of distributed systems, with focus on serverless computing, machine learning systems, and cloud resource management. He published extensively in the premier conferences and journals of his fields. His research has won the Best Paper Runner Up awards of IEEE ICDCS 2021 and USENIX ICAC 2013.
\end{IEEEbiography}

\begin{IEEEbiography}[{\includegraphics[width=1in,height=1.25in,clip,keepaspectratio]{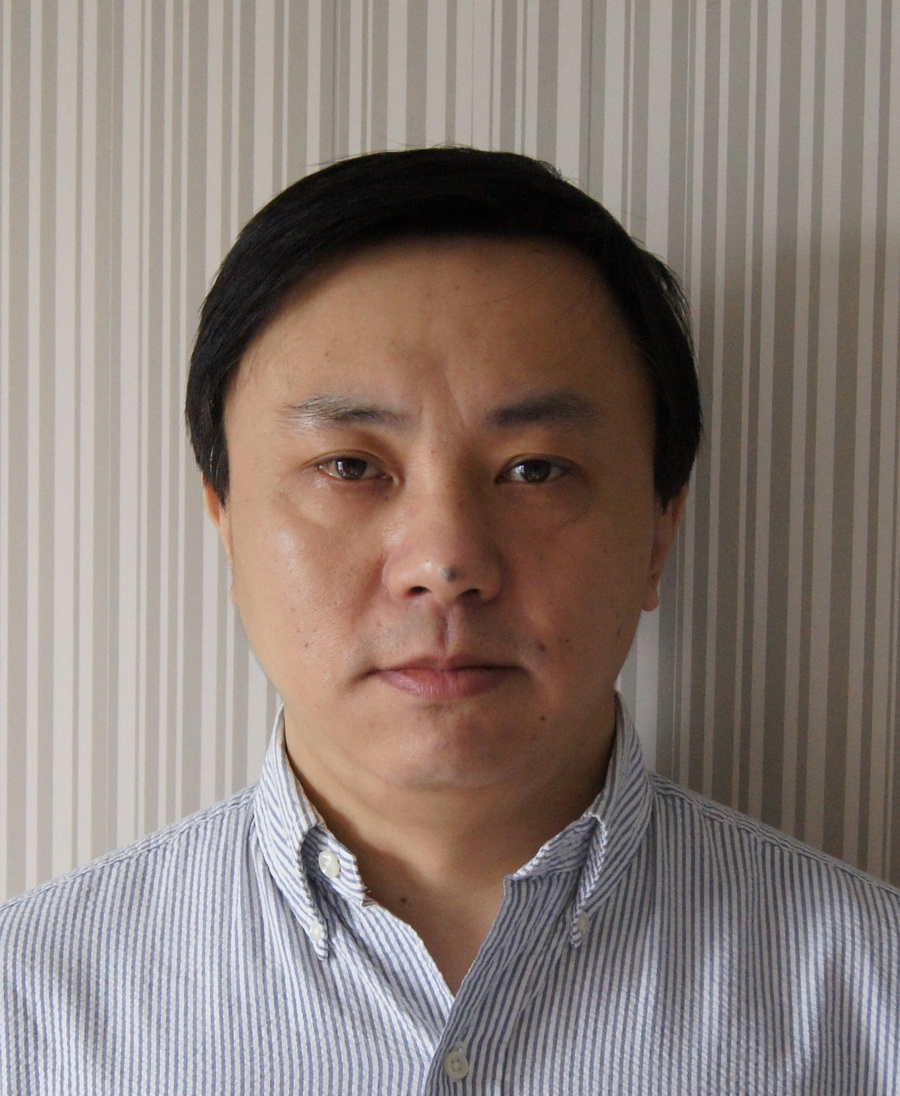}}]{Bo Li} is a Chair Professor in Department of Computer Science and Engineering, Hong Kong University of Science and Technology. He was a Cheung Kong Scholar Visiting Chair Professor in Shanghai Jiao Tong University (2010-2016), and was the Chief Technical Advisor for ChinaCache Corp. (NASDAQ:CCIH), a leading CDN provider. He made pioneering contributions in multimedia communications and the Internet video broadcast, which attracted significant investment and received the Test-of-Time Best Paper Award from IEEE INFOCOM (2015). He received 6 Best Paper Awards from IEEE including IEEE INFOCOM (2021).  He was the Co-TPC Chair for IEEE INFOCOM (2004).
  
He is a Fellow of IEEE. He received his PhD in the ECE Department, University of Massachusetts at Amherst, and his B. Eng. (summa cum laude) in the Computer Science from Tsinghua University, Beijing, China.
\end{IEEEbiography}

\vfill

\newpage

\begin{IEEEbiography}[{\includegraphics[width=1in,height=1.25in,clip,keepaspectratio]{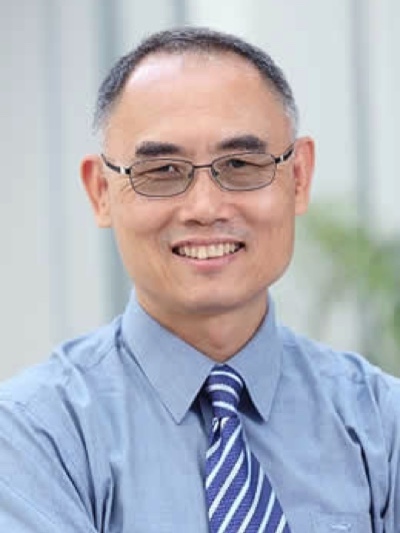}}]{Qiang Yang} received the B.Sc. degree in astrophysics from Peking University, Beijing, China, in 1982, and the Ph.D. degree in computer science and the M.Sc. degree in astrophysics from the University of Maryland at College Park, College Park, MD, USA, in 1985 and 1989, respectively.

He was a Faculty Member with the University of Waterloo, Waterloo, ON, Canada, from 1989 to 1995, and Simon Fraser University, Burnaby, BC, Canada, from 1995 to 2001. He was the Founding Director of the Noah’s Ark Laboratory, Huawei, Hong Kong, from 2012 to 2014, and a Co-Founder of 4Paradigm Corporation, Beijing, an AI platform company. He is currently the Head (Chief AI Officer) with the AI Department, WeBank, Shenzhen, China, and the Chair Professor with the Department of Computer Science and Engineering (CSE), The Hong Kong University of Science and Technology, Hong Kong, where he was the Former Head of the Department of CSE and the Founding Director of the Big Data Institute, Hong Kong, from 2015 to 2018. He has authored several books, including Intelligent Planning (Springer), Crafting Your Research Future (Morgan and Claypool), and Constraint-Based Design Recovery for Software Engineering (Springer). His research interests include artificial intelligence, machine learning, and data mining, with an emphasis on transfer learning, automated planning, federated learning, and case-based reasoning.

Dr. Yang is a fellow of several international societies, including the ACM, AAAI, IAPR, and AAAS. He served as an Executive Council Member for the Association for the Advancement of AI from 2016 to 2020 and the President for the International Joint Conference on AI from 2017 to 2019. He was a recipient of several awards, including the 2004/2005 ACM KDDCUP Championship, the AAAI Innovative AI Applications Award in 2016, and the ACM SIGKDD Distinguished Service Award in 2017. He was the Founding Editor-in-Chief of the ACM Transactions on Intelligent Systems and Technology and the IEEE TRANSACTIONS ON BIG DATA.
\end{IEEEbiography}


% \end{IEEEbiography}


%\newpage



%\vfill


\end{document}

%% file: content/abstract.tex
The increasing demand for privacy-preserving collaborative learning has given rise to a new computing paradigm called federated learning (FL), in which clients collaboratively train a machine learning (ML) model without revealing their private training data. Given an acceptable level of privacy guarantee, the goal of FL is to minimize the \emph{time-to-accuracy} of model training. Compared with distributed ML in data centers, there are four distinct challenges to achieving short time-to-accuracy in FL training, namely the lack of information for optimization, the tradeoff between statistical and system utility, client heterogeneity, and large configuration space. In this paper, we survey recent works in addressing these challenges and present them following a typical training workflow through three phases: client selection, configuration, and reporting. We also review system works including measurement studies and benchmarking tools that aim to support FL developers.

%% file: content/introduction.tex
\IEEEraisesectionheading{\section{Introduction}\label{chap:introduction}}

\IEEEPARstart{B}{uilding} high-quality machine learning (ML) models demands a massive amount of training data. Yet, the communication cost and privacy concerns impinge on the process of collecting large volumes of data from diverse sources. It is not until recently that governments started to regulate the commercial use of data with privacy-preserving legislation (e.g., GDPR~\cite{gdpr}, HIPAA~\cite{hipaa}, and CCPA~\cite{ccpa}). Compliance violations can be costly, with hefty fines up to hundreds of millions of dollars a year~\cite{gdpr-fine, hipaa-fine}. As such, the desire for multiple entities (e.g., mobile devices or large organizations) to collaboratively train a shared model efficiently and privately gives birth to a new ML paradigm called federated learning (FL)~\cite{mcmahan2017communication}. FL promises not to expose the clients' raw data, and has been widely adopted in many industries with applications ranging from mobile devices~\cite{yang2018applied, hard2018federated, ramaswamy2019federated, chen2019federated, assistant, paulik2021federated} to financial management~\cite{ludwig2020ibm, laundering} and medical care~\cite{li2019privacy, oxygen}.

Apart from providing strong privacy guarantees, the key to the success of a federated training system lies in its efficiency. A typical efficiency metric is the \emph{time-to-accuracy}, which is the wall clock time taken to train a model until it reaches the target accuracy. Despite the rich body of work that explored various optimization strategies, there is still plenty of room for further improvement due to the following distinct challenges posed to FL (\cref{chap:background}): (1) the \textit{lack of information for optimization}: the information needed for optimally configuring the system is usually outdated or simply unavailable due to privacy constraints and scaling issues; (2) the \textit{tradeoff between statistical and system utility}: statistical utility (the number of iterations taken to reach a plausible target accuracy) and system utility (the duration of an iteration), the two determining factor for time-to-accuracy, are usually at odds in FL; (3) \textit{client heterogeneity}: clients cannot be treated uniformly due to the intrinsic differences in terms of resources, data, and states; and (4) a \textit{large configuration space}: the operational dimensions for system developers are too many to explore within an acceptable time. Given these challenges, it is worth summarizing existing research efforts to give researchers a holistic view of the lessons learned and to solicit further explorations.

To position existing research attempts in optimizing the time-to-accuracy
performance in FL, we propose a layered approach that categorizes them by the
training phases in which they take effect: selection, configuration, and
reporting (\cref{chap:prior}). For the \textit{selection phase} where the
server chooses clients for participation, there are mainly two lines of
optimization efforts: (1) prioritizing clients either with
high statistical utility or system utility~\cite
{zhang2021client, nishio2019client, wang2020optimizing}, and (2) explicitly considering
both utilities and developing a more informed solution in response to client
dynamics in practice~\cite{chai2020tifl, lai2020oort, kim2021autofl}. 

As for the \textit{configuration phase} where the server sends the global
model to the selected clients with auxiliary configuration information, and
clients perform local training, we sort out four lines of work: (1) the first
two lines advocate mitigating the communication cost by reducing the model
size~\cite{alistarh2017qsgd, wen2017terngrad, bernstein2018signsgd,
wu2018error, karimireddy2019error, jiang2018sketchml,ivkin2019communication,
rothchild2020fetchsgd, spring2019compressing, wangni2017gradient,
aji2017sparse, lin2017deep, stich2018sparsified, shi2019distributed} and
decreasing the synchronization frequency~\cite
{hsieh2017gaia, kamp2018efficient, luping2019cmfl, chen2019communication,
chen2021communication}; (2) the last two lines minimize the computational
overhead by accelerating the training speed in each round~\cite
{anh2019efficient, nguyen2020resource, wang2020towards, li2018federated, diao2020heterofl, ren2020accelerating} and reducing the number of training
rounds~\cite{li2018federated, yu2019linear, yuan2020federated, liu2020accelerating, li2019feddane, acar2021federated, 
karimireddy2020scaffold, karimireddy2021breaking, al2020federated}. 

In terms of the \textit{reporting phase}, we focus on the aggregation and
outline two related optimizations: (1) reducing the aggregation
latency by adopting hierarchical methods~\cite
{liu2020client, abad2020hierarchical, wu2020accelerating} and developing
lightweight privacy-preserving methods~\cite
{so2021turbo, zhang2020batchcrypt, jiang2021flashe}, and (2)
improving the long-term convergence rate with adaptive
optimizers on the server~\cite{hsu2019measuring, wang2019slowmo,
reddi2020adaptive}. For each of the attempted optimizations, our discussion
includes necessary details for readers to understand the motivation,
mechanisms, and major results. In addition, works such as measurement
studies~\cite{yang2021characterizing} and benchmarking tools~\cite
{caldas2018leaf, hu2020oarf, lai2021fedscale, fate, he2020fedml,
beutel2020flower, plato} are indispensable in system research. We also survey the
status quo as a tutorial on FL practice (\cref{chap:concerstone}).

Our work focuses on the system-level efforts made in improving the time-to-accuracy performance for synchronous federated training. We also share some implications derived from the literature and our survey process. It thus differs from existing surveys which mainly focus on ML algorithms and privacy-preserving algorithms. We expect this work to be an initial attempt to bridge the gap of system-oriented surveys in FL literature, as well as soliciting more contributions to related research.

%% file: content/background.tex
%!TEX root=../main.tex
\section{Background, Problem and Challenges}\label{chap:background}

In this section, we give a detailed introduction to the system optimization problem in federated training. We start with a quick primer on the execution workflow of federated training (\cref{sec:background_federated}), followed by the problem statement and the scope of this survey (\cref{sec:background_problem}). We next outline two challenges that make the problem difficult: optimality and practicality (\cref{sec:background_challenge}), which also serves as a summary of criteria for evaluating existing solutions presented thereafter.

\subsection{Federated Training}
\label{sec:background_federated}

Federated learning (FL)~\cite{mcmahan2017communication} has recently emerged
as a new paradigm of collaborative machine learning (ML) that allows multiple
distributed clients (e.g., mobile devices or business organizations) to
collaboratively train or evaluate a model with decentralized data. Compared
to traditional distributed learning in datacenter environments, FL mainly
differs in orchestration, resource constraints, data distribution, and
participation scale~\cite{kairouz2019advances}. At its core, FL keeps private
data on-premises, while using a central server to maintain a global model and
iteratively refine it by aggregating each client's local updates. This design
reduces not only the communication cost but also the privacy risk in
gathering clients' raw data. Owing to its privacy guarantees, FL has found
wide applications in various domains. On mobile devices, Google runs FL to
improve the user experience for Google Keyboard~\cite
{yang2018applied, hard2018federated, ramaswamy2019federated,
chen2019federated} and Assistant~\cite{assistant}, while Apple deploys FL to
evaluate and tune speech recognition models~\cite{paulik2021federated}; in
fintech, both IBM~\cite{ludwig2020ibm} and WeBank~\cite{laundering} utilize
FL to detect financial misconducts; in healthcare, NVIDIA applies FL
to create medical imaging AI~\cite{li2019privacy} and predict patients' needs
for oxygen~\cite{oxygen}.

\begin{figure}[t]
    \centering
    \includegraphics[width=1.0\columnwidth]{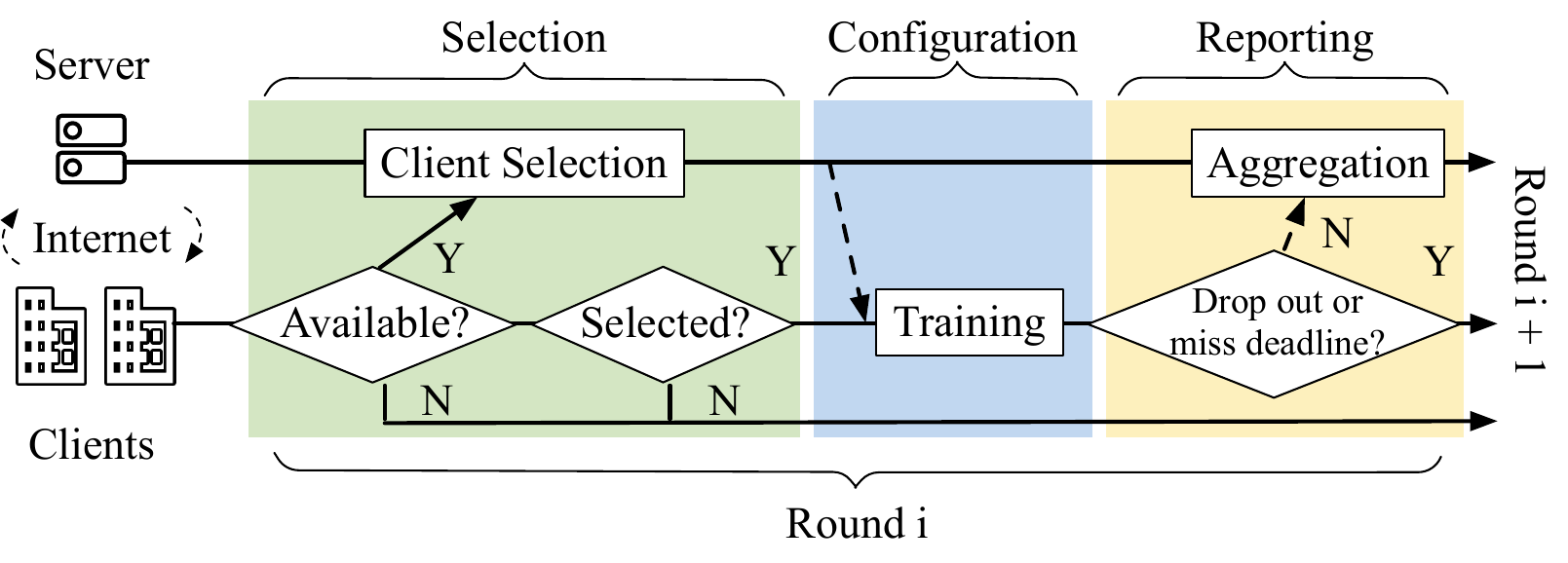}
    \caption{Standard synchronous federated training protocol~\cite{bonawitz2019towards, yang2021characterizing, lai2021fedscale}.}
    \label{fig:fl}
\end{figure}

While both model training and evaluation play important roles in the development of an FL model, they have different criteria in system design. In this survey, we limit the scope to the \textit{training} process, which is the most time-consuming and resource-intensive stage throughout the development of an FL model. Due to its predominance in practice, we focus on the support for the \textit{synchronous} mode, wherein an ML model is trained across a pool of candidate clients in rounds, and in each round, the server needs to wait until a predefined deadline or receiving a sufficient number of clients' updates prior to deriving an aggregated update. In more detail, each round consists of the following three phases (Fig.~\ref{fig:fl}).

\begin{itemize}
    \item \textit{Client Selection}. At the beginning of each round, the server waits for a sufficient number of clients with eligible status (i.e., currently charging and connected to an unmetered network) to check in. The server then selects a subset of them based on certain strategies (e.g., randomly or selectively) for participation, and notifies the others to reconnect later.
    \item \textit{Configuration}. The server next sends the global model status and configuration profiles (e.g., the number of local epochs or the reporting deadline) to each of the selected clients. Based on the instructed configuration, the clients perform local model training independently with their private data.
    \item \textit{Reporting}. The server then waits for the participating clients to report local updates until reaching the predefined deadline. The current round is aborted if no enough clients report in time. Otherwise, the server aggregates the received local updates, uses the aggregate to update the global model status, and concludes the round.
\end{itemize}

\subsection{System Optimization: The Problem\label{sec:background_problem}}

The primary goal of system optimization in federated training is to minimize the end-to-end resource usage of performing a task. The most common metric is the wall clock time, which is typically measured from the very beginning to a certain desirable checkpoint (e.g., convergence or reaching target accuracy). When a metered network is in use (e.g., when clients are on-demand virtual machines in a public cloud), the overall monetary cost becomes another relevant metric that deserves special attention. When uncharged devices are involved, the power consumption should also be considered. Because the cost and energy consumption generally grows linearly as time flies, in this survey, we are particularly interested in reducing the \textit{time-to-accuracy}, i.e., the wall clock time for achieving a preset accuracy target.

\begin{figure*}[t]
    \centering
    \begin{subfigure}[b]{0.19\linewidth}
        \centering
        \includegraphics[width=\columnwidth]{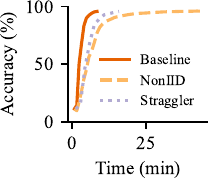}
        \caption{Time to accuracy performance.}
        \label{fig:toy_time_to_acc}
    \end{subfigure}
    \begin{subfigure}[b]{0.19\linewidth}
        \centering
        \includegraphics[width=\columnwidth]{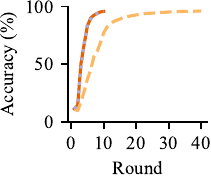}
        \caption{Round to accuracy performance.}
        \label{fig:toy_round_to_acc}
    \end{subfigure}
    \begin{subfigure}[b]{0.19\linewidth }
        \centering
        \includegraphics[width=\columnwidth]{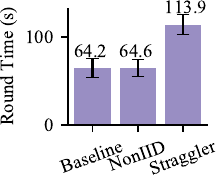}
        \caption{Average round time w/ standard deviation.}
        \label{fig:toy_round_time}
    \end{subfigure}
    \begin{subfigure}[b]{0.19\linewidth}
        \centering
        \includegraphics[width=\columnwidth]{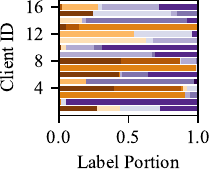}
        \caption{Data distributions of clients in NonIID.}
        \label{fig:toy_noniid}
    \end{subfigure}
    \begin{subfigure}[b]{0.19\linewidth}
        \centering
        \includegraphics[width=\columnwidth]{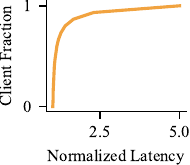}
        \caption{Computational speed of clients in Straggler.}
        \label{fig:toy_straggler}
    \end{subfigure}
    \caption{A case study for illustrating system utility and statistical utility with FL emulation.}         
    \label{fig:toy}
    \end{figure*}

\subsubsection{Statistical Utility and Sytem Utility: A Case Study~\label{sec:background_problem_utility}}

Intuitively, the time-to-accuracy performance of federated training is determined by two factors: the number of rounds taken to reach the accuracy and the average duration of training rounds. As in \cite{lai2020oort}, we regard the former as \textit{statistical utility} and the latter as \textit{system utility} throughout this survey.

To illustrate, we conduct an FL case study where 16 clients collaboratively train the LeNet5~\cite{lecun1989backpropagation} model to classify the images from the MNIST~\cite{deng2012mnist} dataset. To emulate the cross-device environment, each client is run atop an AWC EC2 \texttt{c5.xlarge} instance (4 vCPUs and 8 GB memory) and the network bandwidth is throttled to 21 Mbps to dictate the average mobile connection speed as of 2021~\cite{cisco2021global}. We run FedAvg~\cite{mcmahan2017communication} for model aggregation without loss of generality, atop which we enforce clients' privacy by implementing the SecAgg protocol~\cite{bonawitz2017practical} (also see~\cref{sec:prior_configuration_latency}) which makes the server blind to each client's local updates. Essentially, we design three schemes to compare:

\begin{itemize}
    \item \textit{Baseline}: all clients' data is independent and identically distributed (IID). They also share the same computational speed.
    \item \textit{NonIID}: the same as \textit{Baseline} except that clients' data is non-IID as depicted by Fig.~\ref{fig:toy_noniid}. This is achieved by latent Dirichlet allocation (described in~\cref{sec:cornerstone_benchmarking_datasets}) with the concentration vectors being all 0.1's.
    \item \textit{Straggler}: the same as \textit{Baseline} except that clients’ computation speeds follow the Zipf's distribution (with $\alpha=1.2$, i.e., moderately skewed). As shown in Fig.~\ref{fig:toy_straggler}, the straggler is 5$\times$ slower than the fastest who shares the same speed as clients in \textit{Baseline}.
\end{itemize}

We first study the impact of statistical utility by comparing \textit{Baseline} and \textit{NonIID}. As indicated in Fig.~\ref{fig:toy_round_time}, they proceed a round at the same speed. However, as the learning curves in Fig.~\ref{fig:toy_time_to_acc} dictate, \textit{Baseline} reaches the target accuracy (96.1\%) with only 8.4 min, while \textit{NonIID} takes 43.5 min, which is 5$\times$ slower than \textit{Baseline} does. Referring to the round-to-accuracy performance as in~Fig.~\ref{fig:toy_round_to_acc}, one can know that the reason for this contrast lies in the difference in the numbers of rounds taken to convergence: 10 for \textit{Baseline} while 40 for \textit{NonIID}. Hence, despite having the same system utility, \textit{Baseline} achieves much better time-to-accuracy performance for having greater statistical utility due to more evenly distributed data across clients. Of course, data distribution is uncontrollable in FL practice and developers can barely end up with IID cases as in \textit{Baseline}. This adds up to the challenges of system-level optimizations in FL training, as later discussed in~\cref{sec:background_challenge_optimality}.

Next, we provide a sense of how system utility affects the end-to-end performance by comparing \textit{Baseline} with \textit{Straggler}. As shown in Fig.~\ref{fig:toy_round_to_acc}, both cases converge with the same amount of rounds: 10. However, it takes \textit{Straggler} slightly longer in time (15.8 min, i.e., 1.9$\times$) to reach the target accuracy, as demonstrated in Fig.~\ref{fig:toy_time_to_acc}. By observing Fig.~\ref{fig:toy_round_to_acc} and Fig.~\ref{fig:toy_round_time}, we know that the source of the difference does not stem from the number of rounds used but the making span of each round. As all clients have to proceed at the same speed as the slowest clients in synchronous training, \textit{Straggler} features lower system utility than \textit{Baseline} does due to the presence of slow clients, rendering worse time-to-accuracy. Similarly, the disparity of clients' capabilities is inevitable in FL practice, which also complicates the design of system-level optimizations as mentioned in~\cref{sec:background_challenge_optimality}.

\subsubsection{What is beyond the Scope~\label{sec:background_problem_beyond}}

We emphasize that the solutions discussed in this survey operate at the \textit{system-level}. Therefore, the following directions of work are excluded from discussion, though they could effectively improve the statistical and/or system efficiency in synchronous federated training:

\begin{itemize}
    \item Hardware updates, e.g., adapting programmable switches to enjoy the communication efficiency brought by in-network aggregation~\cite{lao2021atp, amedeo2021scaling}.
    \item Security mechanisms, e.g., employing robust aggregation methods to protect the statistical utility from being impaired by model poisoning attacks~\cite{fang2020local, li2019abnormal}.
    \item Paradigm innovations, e.g., (1) letting clients fit different sets of model parameters by personalization~\cite{smith2017federated, fallah2020personalized, sattler2020clustered, li2021ditto, acar2021debiasing} or models of heterogeneous architectures through knowledge distillation~\cite{li2019fedmd, lin2020ensemble, zhu2021data} to tackle data heterogeneity (a concept mentioned in~\cref{sec:background_challenge_optimality}), (2) allowing clients to exchange data representations for realizing prototype learning~\cite{tan2021fedproto} to reduce communication burdens, or (3) permitting clients to send encoded versions of local datasets to the server to reduce the computational complexity~\cite{prakash2020coded}.
    \item Optimizations on upstream parts of the FL pipeline, e.g., searching for neural architectures that yield better predictive accuracy~\cite{he2020towards, zhu2021federated}.
\end{itemize}

\subsection{What Makes It Hard: The Challenges\label{sec:background_challenge}}

Despite the clear objective, it is non-trivial to work out a feasible solution due to the following two challenges.

\subsubsection{On the Optimality of a Solution\label{sec:background_challenge_optimality}}

First, the information needed in decision-making may be \textit{outdated or even unavailable}. For example, to estimate a client's system utility, it is common to refer to its most recent response latency~\cite{lai2020oort}. However, due to the dynamics over time, such information may not accurately reflect the client's current status. There also exists a cold-start issue, where we are unaware of a client's system capabilities until its first participation. As for estimating a client's statistical utility, the amount of available information is further limited by privacy concerns. According to the recent FL literature, exploratory attacks such as property inference~\cite{melis2019exploiting}, membership inference~\cite{shokri2017membership, song2019auditing}, and data reconstruction~\cite{zhu2019deep, wang2019beyond} can be made possible with model updates. As such, even exposing model updates can discourage clients from participation, let alone inquiring about their data distributions or even raw data~\cite{appledp, shastri2019understanding, shastri2019seven}. Note that the uncertainty in clients' statistical utility and system utility can be accumulated over time.

Even given a holistic view of the environment, the problem remains hard due to the \textit{coupled nature} of statistical utility and system utility. Intuitively, improving system utility is equivalent to minimizing the average resource consumption (e.g., time or bandwidth) per task unit. On the other hand, reducing the resources invested in a task unit inevitably downgrades the quality of the outcome (e.g., statistical utility) as long as no resource is redundant. To exemplify, by constantly picking the fastest clients in client selection, the average duration of each round indeed decreases, whereas the number of rounds taken to target accuracy may be increased as well when other clients' data are under-represented in the global model. Another example can be found in model compression. To improve communication efficiency, a client can send only an important subset of model updates by sparsification~\cite{wangni2017gradient, aji2017sparse, lin2017deep, stich2018sparsified, shi2019distributed}, or a low-bit representation of them by quantization~\cite{alistarh2017qsgd, wen2017terngrad, bernstein2018signsgd, wu2018error, karimireddy2019error}. Although the per-round communication duration can be significantly reduced by adopting a higher compression ratio, the convergence has to take more rounds to occur due to the loss of arithmetic precision.

The problem is further complicated by \textit{client heterogeneity.} Federated training involves tens to potentially millions of clients, each of which intrinsically differs from one another in the following three aspects: 

\begin{itemize}
    \item \textit{Resource Heterogeneity}: due to the variability in hardware specifications and system-level constraints, clients in federated training typically possess different capabilities in computation (CPU/GPU/NPU, memory, and storage), communication (connectivity and bandwidth) and power (battery level and lifespan)~\cite{li2020federated}. These types of heterogeneity complicate the optimization of the overall system utility. For example, merely improving the communication speed does not necessarily lead to shorter end-to-end latency, especially when the straggler is bottlenecked by the computation~\cite{lai2021fedscale}.
    
    \item \textit{Data Heterogeneity}: as the training datasets of clients are typically generated based on their local activities and contexts, they are not IID. More specifically, clients' datasets mainly differ in two aspects\footnote{See a more complete categorization of non-IID scenarios in \cref{sec:cornerstone_benchmarking_datasets}.}: (1) sample quantity (i.e., the number of data samples), and (2) label partition (i.e., the distribution of data labels)~\cite{hsieh2020non}. As a result, not all of them are representative of the population distribution. In case we do not include all the clients in the federation, optimization for statistical utility has to additionally account for such heterogeneity.
    
    \item \textit{State Heterogeneity}: as observed from real-world traces~\cite{yang2021characterizing, lai2021fedscale}, the available slots of mobile device clients vary significantly in temporal distribution due to different user behaviors (e.g., screen locking or battery charging). Therefore, in each round, there can be different sets of candidate clients to choose from, as well as different client drop-out outcomes. On top of the non-IID distribution of clients' data, this type of heterogeneity further complicates statistical utility optimization. Nevertheless, in the cross-silo settings, it may be less of a concern due to the stable and dedicated nature of clients' computing power~\cite{kairouz2019advances, he2020fedml}.
\end{itemize}

Last but not least, it is \textit{infeasible to search through the entire configuration space} for the global optimum. On the one hand, the space is prohibitively large, as a federated training task typically spans $10^1$--$10^6$ users and $10^2$--$10^4$ rounds~\cite{kairouz2019advances}, wherein each phase of a round (\cref{sec:background_federated}) has multiple configurable hyperparameters and alternative policies (e.g., client selection choices in the selection phase, or the number of local steps in configuration phase). On the other hand, most of the online decisions are made on the critical path of the task, meaning that the time spent on working out a solution also counts towards the end-to-end runtime performance, the very objective of the optimization. As a result, it is desirable to be guided by efficient and effective heuristic algorithms, especially balancing the exploration and exploitation efforts made in the configuration space.

\subsubsection{On the Practicality of a Solution\label{sec:background_challenge_practicality}}

\begin{figure*}[t]
    \centering
    \includegraphics[width=0.9\textwidth]{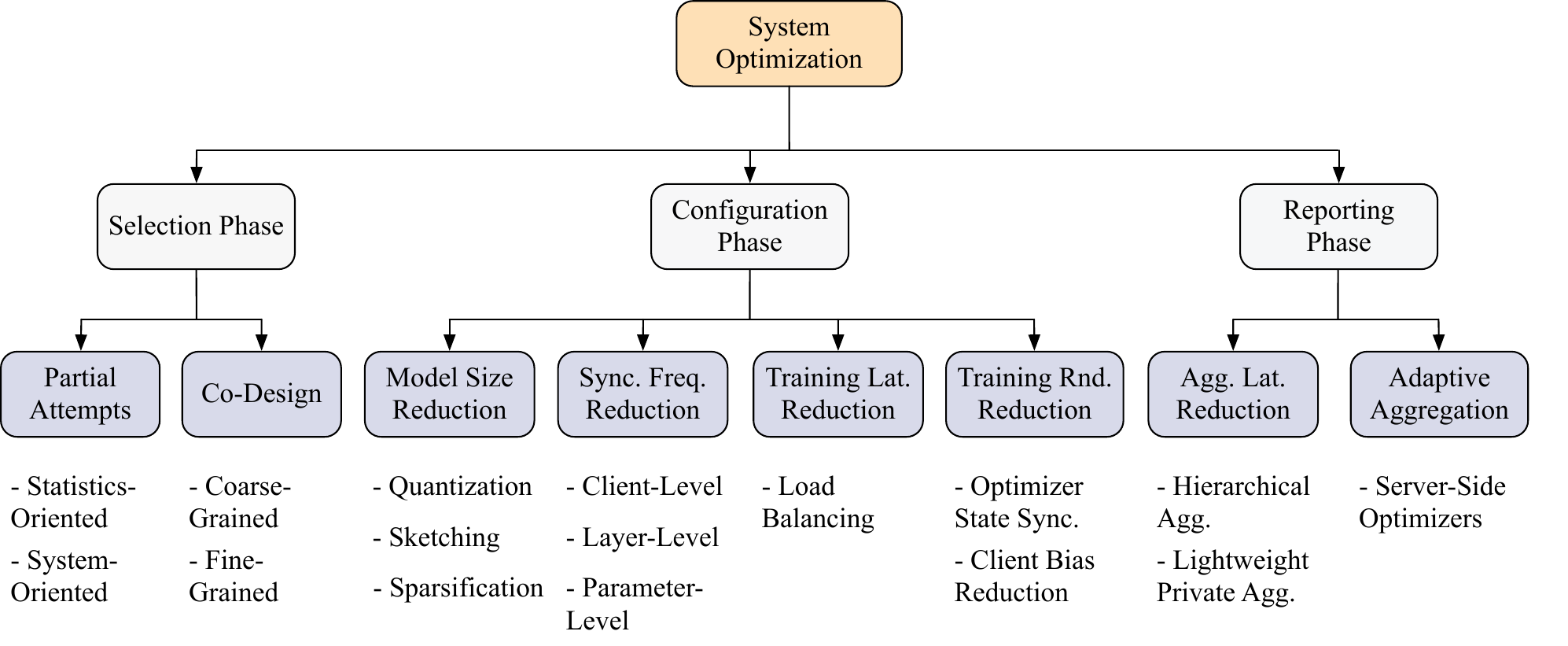}
    \caption{Taxonomy of the approaches discussed in \cref{chap:prior}.}
    \label{fig:opt}
\end{figure*}

Apart from navigating the performance-accuracy-privacy trade-off, the design process of a practical optimization solution should mitigate the accompanying side-effects on other aspects such as the loss of \textit{robustness} to attacks and failures~\cite{kairouz2019advances}. For example, to evaluate the statistical utility of a client, the server may require it to report the loss values generated in local training~\cite{lai2020oort}. However, a malicious or free-rider client may intentionally respond with arbitrary values in the hope of messing with the orchestration or reaping the benefits of the federation without making solid contributions. As such backdoors are introduced by the optimization solution, the developers should take charge of eliminating the undesirable exploitations of these security loopholes. Other possible concerns that may arise as a result of a system optimization solution include but are not limited to \textit{fairness} (e.g., whether participant bias is introduced in the solution), \textit{generality} (e.g., whether the solution applies to diverse tasks), and \textit{ease of deployment} (e.g., whether the solution can be implemented with moderate engineering efforts). In other words, a mature system optimization solution should not only improve the time-to-accuracy performance by enhancing statistical and system utility but also minimize the adverse impacts on other aspects that federated training also values in practice.

%% file: content/prior.tex
%!TEX root=../main.tex

\section{Recent Optimization Approaches\label{chap:prior}}

In the past few years, considerable research efforts have been put into tackling the above challenges for fully unleashing the performance potential of FL training. In this section, we organize them by the training phases, i.e. selection (\cref{sec:prior_selection}), configuration (\cref{sec:prior_configuration}), and reporting (\cref{sec:prior_reporting}), as visualized in Fig.~\ref{fig:opt}.

\subsection{Optimizing the Selection Phase\label{sec:prior_selection}}

Due to the (potentially) large population size and the heterogeneity across clients, the effectiveness of the used participant selection algorithm plays a critical role in the time-to-accuracy performance in federated training. However, the state-of-the-practice system still relies on randomly picking participants~\cite{bonawitz2019towards}, which inevitably leads to waste of resources and suboptimal convergence speed. In response, there is an array of work to guide the selection, which can be roughly categorized by the target utilities that they improve.

\subsubsection{Partial Optimization Attempts\label{sec:prior_selection_either}}

This line of work does not consider the interplay between statistical utility and system utility. Instead, they mainly focus on lifting either utility while leaving the other one ignored or controlled to a limited extent.

\PHM{Statistics-Oriented.} To approach the convergence rate in centralized settings where the data is IID, CSFedAvg~\cite{zhang2021client} advocates that clients with a lower degree of non-IID data should participate more often. To this end, the authors propose weight divergence to capture the non-IID degree of data owned by a client. More precisely, it measures the normalized Euclidean distance between a client's model and the reference model trained by the server with auxiliary IID data. According to the 500-client simulation over CIFAR-10 and Fashion MNIST, CSFedAvg reduces the time-to-accuracy by up to 4.0$\times$ and 2.7$\times$, respectively, compared to random selection.

Besides heuristic methods, some researchers use reinforcement learning (RL) algorithms to learn which clients to select in the presence of data heterogeneity. For example, FAVOR~\cite{wang2020optimizing} seeks to reduce the number of rounds to reach a target accuracy with a deep Q-learning network (DQN)~\cite{mnih2013playing}. To capture each client's statistical characteristics, it takes the low-dimension representations of local models as the RL states. Compared to random selection, FAVOR can reduce the communication rounds by up to 49\% in three image classification tasks. On the other hand, the training overhead for the RL agent may become an obstacle to FAVOR's real-world applications. Specifically, it is reported to take more than 100 episodes to train an agent suited for a specific learning task, which could be prohibitive as each episode corresponds to an entire FL process, i.e., training a global model from scratch for reaching a target accuracy.

\PHM{System-Oriented.} In synchronous training, clients with the lowest system utility bottleneck the speed of a federation round. A straightforward way to bound the time usage is setting a deadline for randomly selected clients' to report updates and ignoring any update submitted after the deadline. To avoid waste of computing resources, FedCS~\cite{nishio2019client} takes a step further by proactively selecting a set of clients whose participation is not likely to miss the deadline according to latency estimation results. As there can be multiple eligible sets, FedCS further favors the solution with the largest scale of participation, which reduces part of the loss in statistical utility. Technically, the whole problem is formalized as a complex combinatorial optimization, and the authors resort to a greedy algorithm for efficient approximation. As indicated in their 1000-client simulation, FedCS outperforms FedLim (modified FedAvg with per-round deadlines imposed) by up to 1.2$\times$ and 1.8$\times$ in the time-to-accuracy when training over the non-IID CIFAR-10 and Fashion MNIST datasets, respectively.

\subsubsection{Co-Optimizing Statistical/System Utility\label{sec:prior_selection_both}}

Given the coupled nature of clients' system utility and statistical utility, it is more desirable to navigate the sweet point of jointly maximizing both of them.

\PHM{Coarse-Grained.} TiFL~\cite{chai2020tifl} first considers increasing the system utility. To that end, it divides clients into different tiers based on the observed runtime performance, and at each round only selects clients from the same tier for mitigating the waste of resources due to idle waiting for stragglers. To reduce the average iteration span, it also limits the number of times a (slow) tier can be selected. On top of that, the statistical utility is respected by prioritizing tiers with lower testing accuracy whenever there is more than one electable tier. Compared with FedCS, TiFL bears some resemblance in limiting the participation of less capable clients, while being more aware of the statistical utility. As reported in a 50-client cluster with 5 client tiers, TiFL achieves an improvement over random selection by up to 3$\times$ speedup in overall training time and by 6\% in accuracy.

\PHM{Fine-Grained.}  Compared to TiFL, Oort~\cite{lai2020oort} reconciles the demand for enhancing both system utility and statistical utility in finer granularity. Specifically, it associates each client with a continuous score and prioritizes those clients with higher scores. The score is meant to be a principled measurement of both the statistical utility (determined by the training loss) and the system utility (estimated from historical response latency). As some components of the score cannot be known in advance until the corresponding client's first participation, or cannot be guaranteed to be stable due to the client's runtime dynamics, the score estimation process is modeled as a Multi-Armed Bandit (MAB) problem and tackled by the Upper Confidence Bound (UCB) algorithm~\cite{auer2002finite}. Apart from the scoring backbone, Oort also aims to address other practical issues like staleness and robustness. Oort was evaluated on a 1300-client GPU cluster with realistic datasets and simulation on the client heterogeneity. Compared to random selection, it reduces the training time by up to 14$\times$ and improves model accuracy by up to 9.8\%.

Besides the UCB algorithm, researchers also experiment with more sophisticated RL methods to cherrypick participants. Notably, AutoFL~\cite{kim2021autofl} learns to select participants and execution targets for each individual based on the Q-Learning algorithm~\cite{watkins1992q}. To achieve efficient FL execution, it identifies RL states that are critical to energy efficiency, convergence time, and accuracy. It also defines RL rewards that track the energy consumption of clients and model accuracy. Compared to random selection, AutoFL is reported to improve the average FL energy efficiency by up to 4.3$\times$, while also exhibiting better training accuracy. Similar to FAVOR (mentioned in \cref{sec:prior_selection_either}), however, training the Q-Learning model to converge needs multiple episodes, each of which corresponds to an entire FL training process. Its practicality could thus be challenged whenever offline training is infeasible or costly.

\subsection{Optimizing the Configuration Phase\label{sec:prior_configuration}}

In the configuration phase, there are mainly two processes that are responsible for the time-to-accuracy performance. One is downlink (i.e., server-to-client) model transmission and the other is local model training. Thus, both aspects can be investigated for system optimization. As for communication overhead reduction, one can reduce the size of model updates (\cref{sec:prior_configuration_size}) and decrease the synchronization frequency (\cref{sec:prior_configuration_freq}). To lower computational overhead, one can shorten the training latency by balancing the workload across clients (\cref{sec:prior_configuration_latency}), as well as reducing the number of rounds taken to converge by adopting heterogeneity-aware training algorithms (\cref{sec:prior_configuration_round}). As the uplink model submission (i.e., client-to-server) that takes place in the reporting phase shares the same operational space as the downlink one, we combine the discussion on both of them in this section for brevity.

\subsubsection{Model Update Size Reduction\label{sec:prior_configuration_size}}

Prior arts of model update size reduction mainly fall into three camps: quantization, sketching, and sparsification. Ahead of the emergence of FL, the exploration of these directions has already been initiated in the context of traditional distributed learning. While their communication merits are mostly reproducible in FL, they also face new challenges due to the privacy regulations and client heterogeneity, which we will also point out hereafter.

\PHM{Quantization.} Quantization converts each scalar in a model update to its low-bit representation which takes up less space. While quantization has already gained its fame in traditional distributed learning and we refer the readers to dedicated surveys like~\cite{tang2020communication} for more details, here we only introduce the most representative work. As the first quantization work in model training with rigorous convergence proof, QSGD~\cite{alistarh2017qsgd} performs unbiased quantization with standard random dithering, a technique borrowed from image processing. Since its birth, related works have been emerging with more aggressive quantization bit-widths and more appealing empirical performance. For example, TernGrad~\cite{wen2017terngrad} advocates using only ternary values (0, $\pm$1) in the uplink direction, while signSGD~\cite{bernstein2018signsgd} can use only binary signs ($\pm$) in both uplink and downlink communication. It is worth mentioning that a popular technique in tackling the precision loss brought by quantization is error feedback, whose basic idea is to accumulate the previous quantization errors and compensate for them in the current round. Leveraging this technique, ECQ-SGD~\cite{wu2018error} performs consistently better than QSD in terms of both convergence speed and accuracy, while EF-SGD~\cite{karimireddy2019error} also achieves a narrower generalization gap from centralized training compared to signSGD.

Despite their generality, there are some practical concerns about applying these general quantization strategies to FL due to privacy constraints and client heterogeneity. For example, determining the clipping threshold for quantization needs to exploit the knowledge about its inputs (i.e., local model updates) for reducing the induced error as in dACIQ~\cite{banner2018post}. However, an FL client does neither possess a priori knowledge of others' model updates nor should it request these sensitive values. To work out a globally applicable clipping threshold, we may need to share some less sensitive information (e.g., the maximum and minimum values in local updates) across clients for threshold estimation as in BatchCrypt~\cite{zhang2020batchcrypt}. Still, whether such a circumvention guarantees robust estimation and immunity to privacy attacks remains an open question.

\PHM{Sketching.} Existing quantization approaches assume the input values follow a certain distribution (e.g., uniform or bell-shaped), which may not always be the case in model updates~\cite{jiang2018sketchml}. To be more general, some researchers introduce sketching methods where some memory-saving data structures are used to approximate the exact distribution of model update values in a single processing pass over the values. For example, SketchML~\cite{jiang2018sketchml} utilizes a quantile sketch method to generate a non-uniform mapping from gradient values to low-bit integers. SketchML achieves empirical success such as decreasing the gradient size by around 7$\times$ and is the first effort to introduce sketching for compressing model updates in ML training. Similar to quantization, sketch algorithms can also make use of error feedback techniques to efficiently amend the errors induced by the approximation, as demonstrated in SketchedSGD~\cite{ivkin2019communication} and FetchSGD~\cite{rothchild2020fetchsgd}. There is also sketching practice that compresses auxiliary variables apart from model updates, such as sketching clients' momenta and per-coordinate learning rates as in~\cite{spring2019compressing}.

\PHM{Sparsification.} While quantization and sketching operate at the precision level in terms of model size reduction, sparsification operates at the coordinate level. Specifically, sparsification allows each client to transmit only a sparse subset of its model updates, while the rest are accumulated and incorporated into future training. Technically, the sparsified gradient is obtained by first performing element-wise multiplication on the original gradient with a certain 0/1 mask and then discarding zero elements. The mask is typically randomly generated as in~\cite{wangni2017gradient}, while another commonly used variant is the top $s$\% scheme where 1 is given to the coordinates that rank top $s$\% in absolute magnitude and 0 otherwise~\cite{aji2017sparse, lin2017deep, stich2018sparsified, shi2019distributed}. The top $s$\% method is reported to reduce network traffic by up to three orders of magnitude, while still preserving model quality~\cite{aji2017sparse, lin2017deep}.
    
While similar cost savings are transferable to plaintext FL, it is unclear whether sparsification can be further compatible with cryptographic techniques that are widely adopted for privacy enforcement in FL. For example, apart from the uplink model updates, it is also desirable to sparsify the downlink global update for fully releasing the potential for communication improvement. However, implementing the downlink sparsification may not be feasible when the server is not aware of the plaintext values of the aggregated update as a result of the applied Secure Multi-Party Computation (SMPC)~\cite{zhang2020privcoll, bonawitz2017practical, so2021turbo, mohassel2017secureml} or Homomorphic Encryption (HE)~\cite{liu2019secure, zhang2020batchcrypt, smart2010fully, smart2014fully} techniques.

It is noteworthy that as quantization (or sketching) and sparsification are orthogonal to each other, they can be combined to reap the most benefits in terms of model size reduction~\cite{konevcny2016federated, caldas2018expanding}.

\subsubsection{Synchronization Frequency Reduction\label{sec:prior_configuration_freq}}

At its core, the reduction in the synchronization frequency is achieved by identifying and precluding redundant synchronization efforts. This can be operated at different granularities ranging from clients, layers to individual parameters in a model update.

\PHM{Client-Level.} The importance of an entire model update is usually measured by some numerical features. In the most intuitive form, a model update in Gaia~\cite{hsieh2017gaia} is considered significant if its magnitude relative to the current value $\mid \frac{Update}{Value} \mid$ exceeds a specific threshold such as 1\%. While the magnitude may serve as a good indicator for how data center learning performs, it does not work in FL where determining an appropriate threshold is hard due to clients' heterogeneity. As such, some researchers propose to involve the comparison with some reference points for more robust measurement of the importance. For example,~\cite{kamp2018efficient} tracks the Euclidean distance between the local model and a reference model, while CMFL~\cite{luping2019cmfl} focuses on the number of coordinates with the same sign in the local model and the most recent global model.

\PHM{Layer-Level.} Apart from considering a model update as a whole, another line of work tries to reduce the synchronization frequency on a layer basis. A representative work done in this direction is TWAFL~\cite{chen2019communication} where the model aggregation is conducted layer-wise. As observations made in deep neural network (DNN) fine-tuning~\cite{yosinski2014transferable}, shallow layers in a DNN learn general features across different datasets while deep layers learn ad hoc ones. TWAFL hence proposes to update shallow layers more frequently than deep ones as they are more responsible for the overall quality of the global model.

\PHM{Parameter-Level.} Some industrial practitioners also consider whether to synchronize for each round at the level of individual parameters. Noticing that each parameter usually evolves in a transient-then-stable manner, i.e., it first varies drastically and then settles down around a certain value with slight oscillation, APF~\cite{chen2021communication} proposes to stop synchronizing those parameters whose evolution has reached a stationary phase.

\subsubsection{Training Latency Reduction\label{sec:prior_configuration_latency}}

A client's training latency is determined by both its computational workload and resource capabilities. While the latter cannot be altered, the former still leaves room for optimization innovations. We discuss one major line of such efforts.

\PHM{Load Balancing.} Given the variations in computing power and data volume, clients may not finish the training process at the same time. To mitigate the resulting straggler effects, \cite{anh2019efficient, nguyen2020resource} suggest balancing the amount of training data across clients. Specifically, they turn to RL techniques for determining the optimal number of data units used in a communication round for each participant, intending to minimize the time and energy consumption and maximize the volume of involved data. While these solutions can reduce the round latency, the number of rounds needed for convergence does not necessarily remain unchanged because they do not account for data heterogeneity and thus the contribution of slow clients with important data could be throttled. To jointly consider heterogeneous system speed and data distribution during load balancing, \cite{wang2020towards} carefully formulates an optimization problem where each client is associated with a weight that diminishes exponentially with the disparity between its number of labels and that of the population. Compared to even data allocation, this work manages to reduce the end-to-end time-to-accuracy.
    
Load balancing can also be achieved by varying the number of optimization steps or the complexity of local models. For example, FedProx~\cite{li2018federated} balances the system load across clients by formulating an inexact learning problem that allows variable steps of local solvers. On the other hand, HeteroFL~\cite{diao2020heterofl} assigns sub-models with different widths of hidden channels to clients so that clients with fewer capabilities can train smaller sub-models. All sub-models share the same model architecture, and thus normal model aggregation is still possible. The authors empirically show that the quality of the global model trained with heterogeneous sub-models is comparable to that trained with full-sized local models.

Apart from the computational load, it is sometimes also beneficial to balance the communication load across clients, especially when the network conditions are complicated as in wireless connections. For example, targeting a mobile edge computing (MEC) scenario where Time Division Multiple Access (TDMA) is implemented, \cite{ren2020accelerating} optimizes both the data batch size and uplink/downlink frame time slots for each client to achieve the maximum learning efficiency. In addition to coping with CPU computing, the authors further extend the optimization problem to the scenario where devices are equipped with GPUs for training.

\subsubsection{Training Round Reduction\label{sec:prior_configuration_round}}

In the settings with heterogeneous data, more local computation in a communication round does not necessarily lead to fewer numbers of rounds for reaching satisfying optima~\cite{al2020federated}. Thus, adopting heterogeneity-aware techniques such as adaptive optimization and bias reduction can help guarantee the convergence speed in FL practice.

\PHM{Optimizer State Synchronization.} It is common practice for first-order moment optimization to apply momentum to dampen oscillations~\cite{sutskever2013importance, zhang2020adaptive}. However, in the federated settings, if the clients' momenta are separately updated, they may deviate from each other due to non-IID data distributions. Thus, there are researchers proposing optimizer state synchronization frameworks where clients' optimizer states are synchronized by the server periodically. PR-SGD-Momentum~\cite{yu2019linear} is aligned with this direction and also gives proof of the linear speedup of convergence w.r.t. the number of workers. FedAC~\cite{yuan2020federated} also applies momentum at clients with periodic synchronization, while it is proven to obtain the same linear speedup property with asymptotically fewer rounds of synchronization. MFL~\cite{liu2020accelerating} is another similar idea with theoretical guarantees, but it focuses on accelerating deterministic gradient descent (DGD) instead of SGD, unlike the previous two studies.

\PHM{Client Bias Reduction.} Due to data heterogeneity, clients' model updates can be biased towards the minima of local objectives, known as ``client drift'' in the literature~\cite{wang2021field}, which hinders the convergence of the global model. To reduce the variance across clients, a natural idea is to regularize local objective functions for minimizing the drift. For example, assuming bounded dissimilarity between local functions, FedProx~\cite{li2018federated} limits the Euclidean distance between local models and the global one by regularizing the square of the distance. FedDANE~\cite{li2019feddane} uses the same regularization term, while it formulates the other part of the objective function following the Distributed Approximate NEwton (DANE) method~\cite{shamir2014communication} in classic distributed optimization. Despite having a similar theoretical convergence rate as FedProx, FedDANE underperforms FedProx in practice where the data heterogeneity is high and the participation rate is low, suggesting a discrepancy between theory and practice which needs further investigation. The objective function in FedDyn~\cite{acar2021federated} also considers the combination of a linear term and an L2 regularizer, with a different linear term which is formulated to align the empirical loss surfaces of clients. In theory, FedDyn ensures that the consensus point of model convergence across clients will be consistent with the global stationary solution as long as local models converge, regardless of the degree of data heterogeneity.

Apart from regularizing local objectives, variants in clients' local updates can also be reduced by leveraging the idea of control variates borrowed from the convex optimization literature. For example, in SCAFFOLD~\cite{karimireddy2020scaffold}, each of the clients and the server maintains a control variate, and at each local step, a client de-biases its local updates with two control variates: one of its own and the other broadcast by the server. SCAFFOLD converges provably faster than FedAvg~\cite{mcmahan2017communication} without any assumption made on the client selection or data heterogeneity. MIME~\cite{karimireddy2021breaking} considers a similar idea but makes a different choice on the specific definition of control variates.

While the use of control variates requires persistent client states, there exists another line of work that works for stateless clients: posterior averaging. Instead of approaching FL as optimization, this line of work formulates the problem as a posterior inference one. Compared to traditional federated optimization, posterior inference can benefit from an increased amount of local computation without risking stagnating at inferior optima. FedPA~\cite{al2020federated} instantiates this idea with an efficient algorithm to conduct federated posterior inference with linear computation and communication costs.

\subsection{Optimizing the Reporting Phase\label{sec:prior_reporting}}

In this phase, the operation room for system optimization is limited to either model uploading or model aggregation. As the former has already been combined in the last section, we hereafter focus on optimizing the aggregation process with two main directions explored in the literature: (1) directly reducing the aggregation latency at each round (\cref{sec:prior_reporting_round}), and (2) expediting the convergence rate in the long run through conducting adaptive aggregation (\cref{sec:prior_reporting_end}).

\subsubsection{Aggregation Latency Reduction\label{sec:prior_reporting_round}}

Compared to local training in the configuration phase, model aggregation involves less intensive computation. However, its latency can still be salient because (1) large-scale participation can put pressure on the communication, and (2) the deployment of security methods can complicate the computation. We hereafter introduce the respective optimization efforts in the literature.

\PHM{Hierarchical Aggregation.} The downsides of the traditional two-layer (server-client) system include (1) instability: the network bandwidth may be slow or unpredictable, especially in public networks and/or under geo-distributed settings; (2) risk of scalability: the cloud server may suffer from network congestion when concurrently receiving too many local updates; (3) heterogeneity: the straggler effects could be exacerbated by imbalanced network bandwidth. 
    
To address these issues, some researchers resort to a hierarchical design of model aggregation by introducing an extra level of edge servers, each of which is typically responsible for a small number of clients with proximity. For instance, in HierFAVG~\cite{liu2020client}, after a fixed number of local updates on clients, each edge server aggregates its own clients' models. Subsequently, after another fixed interval of edge aggregation, the cloud server aggregates all the edge servers' models. It is proven that HierFAVG still guarantees convergence, and empirical studies with synthetic FL datasets show that it reduces the time-to-accuracy by up to 2.7$\times$ in a simulated cloud-edge-client environment. A concurrent work HFL~\cite{abad2020hierarchical} also considers a similar design, while it does not attach theoretical analysis on its convergence behaviors. HybridFL~\cite{wu2020accelerating} further extends this primary design with two ideas: (1) quota-triggered edge-level aggregation: edge nodes stop waiting for more local updates once receiving a sufficient number of them; and (2) immediate cloud aggregation: cloud-level aggregation is conducted right after the edge-level one is completed. This decouples each pair of interactions (i.e., cloud-edge and edge-client), thereby further mitigating the impact of client drop-out and stragglers.

Despite changing the aggregation rules, the hierarchical designs mentioned above still focus on establishing the convergence to a single global model, which does not deviate from the learning paradigm discussed throughout this survey. On the other hand, there also exist other formulations of collaborative learning such as clustered FL where clients are assigned to different groups and aggregation takes place within a group \cite{mansour2020three, ghosh2020efficient, xie2021multi}. As they aim to build personalized models for each group of clients, they are considered paradigm innovations to standard FL and thus fall out of the scope of this survey (\cref{sec:background_problem}).

\PHM{Lightweight Privacy-Preserving Aggregation.} As mentioned in \cref{sec:background_challenge_optimality}, uploading model updates in the clear may be vulnerable to exploratory attacks which plague clients' privacy. Therefore, model aggregation is preferably safeguarded by cryptographic techniques, which inevitably induces extra computation and communication overhead. 

One line of this work leverages secure multiparty computation (SMPC). One of the most impactful works is Google's secure aggregation (SecAgg) protocol, where pseudorandom masks are used for data confidentiality and Shamir's secret sharing scheme~\cite{shamir1979share} is used to accommodate client dropout~\cite{bonawitz2017practical}. It has provable rigorous privacy guarantees under both passive and active adversary models. Given its quadratic communication overhead during secret sharing, SecAgg has inspired many other SMPC protocols with improved performance. For example, SecAgg+~\cite{bell2020secure} optimizes SecAgg in both communication and computation by replacing the complete graph with a random sparse one of logarithmic degree. TurboAgg~\cite{so2021turbo}, on the other hand, attempts to optimize SecAgg through the use of multi-group circular topology, additive secret sharing, and Lagrange coding. Compared to SecAgg+, TurboAgg incurs a slightly higher overhead in the asymptotic sense for both communication and computation, and only focuses on tackling honest-but-curious adversaries. While all of the above schemes rely on Shamir's scheme for sharing clients' secrets, there also exist other secret sharing techniques that offer different trade-offs across performance, privacy, and dropout resilience, e.g., encoding using Maximum Distance Separable matrices as in LightSecAgg~\cite{yang2021lightsecagg}, or Fast Fourier Transform as in FastSecAgg~\cite{kadhe2020fastsecagg}.
    
Besides SA, Homomorphic Encryption (HE) is another commonly used privacy-preserving aggregation technique that comes with prohibitively high message inflation and runtime overhead. In response, BatchCrypt~\cite{zhang2020batchcrypt} implements an end-to-end solution for batching multiple plaintexts into one large plaintext so that HE-related operations can be performed in a data-parallel manner. BatchCrypt is shown to speed up the training by 23$\times$-93$\times$ compared to plain Paillier~\cite{paillier1999public} (a prevalent variant of HEs), but it still leaves the message inflation suboptimal and is incompatible with top s\% sparsification approaches~\cite{jiang2021flashe}. Instead of optimizing traditional HE schemes by batching, FLASHE~\cite{jiang2021flashe} proposes a lightweight HE scheme that is tailored for cross-silo FL. It induces negligible ($\leq6\%$) computational overhead and no network communication overhead compared to plaintext FL for staying symmetric. Its performance is also optimized when interacting with sparsification techniques.

\subsubsection{Adaptive Aggregation\label{sec:prior_reporting_end}}

In FedAvg~\cite{mcmahan2017communication}, the de facto standard aggregation method, local model updates simply get weighted by the corresponding numbers of training samples and then added up. While it guarantees convergence when even dealing with non-convex empirical risk functions in IID data settings~\cite{yu2019parallel, karimireddy2020scaffold}, it is observed to yield unstable convergence behavior or even divergence when it faces models trained with arbitrarily non-IID data. There are thus rising interests on whether the aggregation can be more adaptive w.r.t. the data heterogeneity across clients.

\PHM{Server-Side Optimizers.} Other than accelerating convergence with local momentum (\cref{sec:prior_configuration_round}), there are also exploration efforts on server-side momentum. As there is originally no optimizer at the server in FL, these methods first need to generalize the existing aggregation algorithm. Specifically, at each round, instead of collecting local model weights, the server instead collects their changes and treats these changes as the ``pseudo-gradient'' for the server, which the server can use to update the global model with adaptive optimizers. FedAvgM~\cite{hsu2019measuring} initiates the empirical studies with the simplest form of momentum applied at the server, while SlowMo~\cite{wang2019slowmo} independently proposes a similar scheme and also attaches the theoretical analysis for its convergence behaviors. A recent work~\cite{reddi2020adaptive} makes more sophisticated use of momentum such as adopting AdaGrad~\cite{duchi2011adaptive}, Adam~\cite{kingma2015adam} and YOGI~\cite{reddi2018adaptive} optimizers (which correspond to FedAdaGrad, FedAdam, and FedYOGI, respectively). It is shown that FedYOGI consistently outperforms FedAvgM in terms of validation performance for both sparse- and dense-gradient FL tasks. Server-side momentum methods feature no need for persistent states or computation burdens at the client end, making them well suited to cross-device scenarios.

%% file: content/benchmark.tex
%!TEX root=../main.tex
\section{Measurement and Benchmarking Tools\label{chap:concerstone}}

\begin{figure}[t]
    \centering
    \includegraphics[width=0.85\columnwidth]{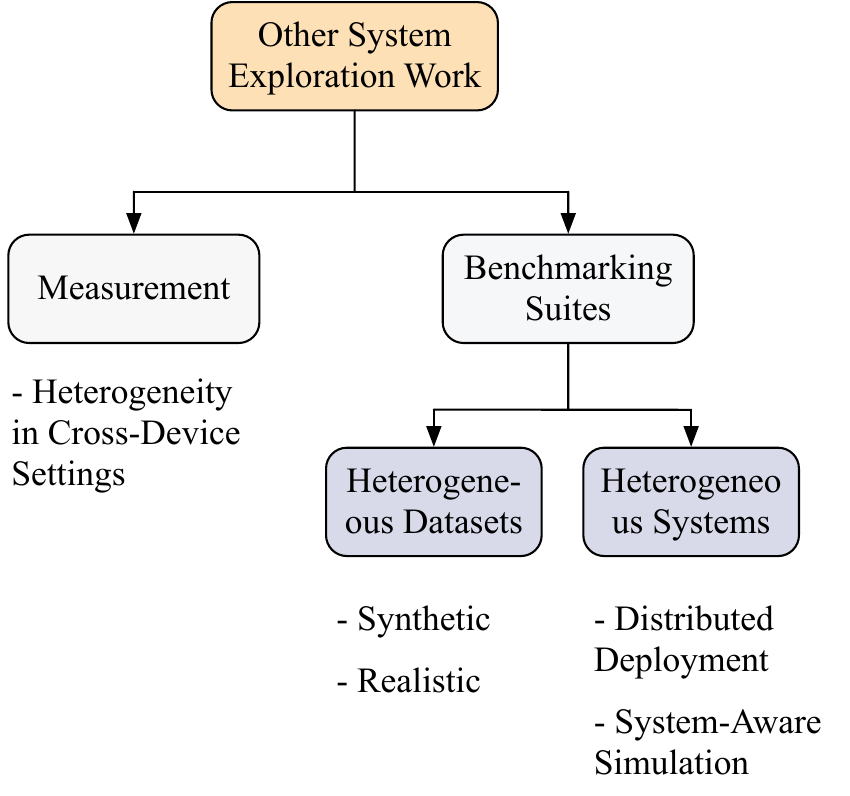}
    \caption{Taxonomy of the work introduced in \cref{chap:concerstone}.}
    \label{fig:cornerstone}
\end{figure}

Aside from innovating optimization solutions, there are also researchers contributing with cornerstone works that benefit the community with informative insights from systematical measurement studies (\cref{sec:cornerstone_measurement}) and grounding benchmarking tools (\cref{sec:cornerstone_benchmarking}), as visualized in Fig.~\ref{fig:cornerstone}.

\subsection{Measurement-Based Research\label{sec:cornerstone_measurement}}

Due to the complicated interplay between statistical utility and system utility, there are a few measurement studies which dedicate to conducting thorough investigations and providing actionable implications for interested researchers.

\begin{itemize}
    \item \textit{FLASH}\footnote{We name it after the corresponding Github repository handle~\cite{flash}.}~\cite{yang2021characterizing}: FLASH particularly studies the impacts that heterogeneity has on both the statistical utility and system utility. This work contributes to the community mainly with three aspects: (1) it establishes a novel large-scale dataset that reflects the system and state heterogeneity among 136,000 real-world smartphones; (2) it implements and open-sources an FL platform that provides a heterogeneity-aware environment for experiments; (3) it reports comprehensive findings that stem from the authors' extensive measurements atop the platform.
    
    Among the authors' numerous implications, there are two fresh findings at the time when the study is written. First, gradient compression methods (e.g., Gradient Dropping~\cite{aji2017sparse} and SignSGD~\cite{bernstein2018signsgd}) can hardly shorten the convergence time under heterogeneous cross-device settings. Second, advanced aggregation algorithms that overlook some aspects of heterogeneity will be less effective in realistic settings. Readers can refer to the paper for more details.
\end{itemize}

\subsection{Benchmarking Suites\label{sec:cornerstone_benchmarking}}

Realistic benchmark suites are necessary for enabling fair, insightful, and reproducible evaluation of the effectiveness of system optimizations. As the time-to-accuracy performance relies on both the statistical utility and system utility (\cref{sec:background_problem}), in the following summary of existing benchmarking tools, we aim to cover diverse aspects of simulating practical FL: data characteristics, client capabilities, and availability.

\subsubsection{Training Datasets\label{sec:cornerstone_benchmarking_datasets}}

There are two prevalent categories of training datasets. One line of work is \textit{derived from conventional ML datasets} (e.g., \texttt{CIFAR}~\cite{krizhevsky2009learning}, \texttt{MNIST}~\cite{wan2013regularization}, and \texttt{Fashion-MNIST}~\cite{xiao2017fashion}). To synthesize the non-IID nature as in real FL scenarios, the data partitions in these datasets are typically formed by restricting the number of data classes each client has (e.g., partitioning by shard-based methods as in~\cite{mcmahan2017communication} or latent Dirichlet allocation (LDA) processes as in~\cite{hsu2019measuring, reddi2020adaptive, al2020federated, acar2021federated}). Although the data generated in such a way are indeed non-IID, they may not perfectly represent the real-world characteristics. For instance, besides the label distribution skew, in reality, non-IID data may also involve feature distribution skew (e.g., same words with different stroke widths), same labels with different features (e.g., images of clothing vary due to regional differences) and same features with different labels (e.g., the same context mapped to different next words due to personal habits)~\cite{hsieh2020non, kairouz2019advances}.

In contrast, the other type of datasets is \textit{collected in real distributed scenarios} and thus naturally captures the FL features. We briefly introduce existing open-source attempts in curating such datasets as follows.

\begin{itemize}
    \item \textit{LEAF}~\cite{caldas2018leaf}: LEAF is an actively maintained project, which currently consists of 6 datasets spanning multiple applications such as computer vision (CV) (\texttt{FEMNIST} and \texttt{CelebA}) and natural language processing (NLP) (\texttt{Shakespeare}, \texttt{Reddit}, and \texttt{Sentiment140}). Each dataset is generally formed by splitting the corresponding public dataset by the original contributors of the data samples. In other words, the non-IID nature comes from the unique behavior style of each contributor.
    \item \textit{FedScale}~\cite{lai2021fedscale}: Similar to LEAF, FedScale also collects realistic datasets and partitions them by unique client identification. FedScale currently includes 18 datasets (including \texttt{iNature}, \texttt{OpenImage}, \texttt{Google Landmark} etc.) which span 10 FL tasks. Apart from the comprehensive coverage of tasks, FedScale has further made four contributions to the community: (1) it has the training, validation, and testing set well established; (2) it streamlines different datasets into a unified format; (3) it accounts for various participation scale from hundreds to millions of clients; and (4) it provides handy APIs for users to customize their datasets.
    \item \textit{OARF}~\cite{hu2020oarf}: for a specific FL task, OARF assembles real-world datasets from different sources to realize data heterogeneity. For example, for sentiment analysis, it combines both the \texttt{IMDB Movie Review} and \texttt{Amazon Moview Review} datasets. In training, datasets belonging to different sources are distributed to different parties. As such, the data are split in a dataset-wise manner instead of a sample-wise one. OARF currently covers 9 tasks in CV and NLP.
\end{itemize}

Besides the above systematic collection, there are also other separately maintained realistic datasets, such as \texttt{Stackoverflow}~\cite{stackoverflow} and \texttt{PERSONA-CHAT}~\cite{zhang2018personalizing}.

\subsubsection{Production Systems and Simulation Platforms\label{sec:cornerstone_benchmarking_platforms}}

In addition to setting up data heterogeneity, we also need to incorporate system heterogeneity in realistic benchmarking. The most straightforward way to study FL designs with system utility borne in mind is to deploy in \textit{production-oriented systems}. Such systems not only embed the ML backbone but also address practical problems like authentication, communication, encryption, and deployment to physical distributed environments. We sketch the open-source representatives.

\begin{itemize}
    \item \textit{FATE}~\cite{fate}: FATE is an FL framework that can be deployed in distributed environments. In addition to its flexible ML pipeline, FATE also features several aspects that further facilitate the research on various goals of practical FL: (1) it supports privacy-preserving computation by implementing cryptographic algorithms such as the Diffie-Hellman key agreement~\cite{diffie1976new} and homomorphic encryption~\cite{paillier1999public}; (2) it covers different training architectures including horizontal FL, vertical FL, and federated transfer learning; and (3) it allows a certain degree of customization on the FL pipeline such as the aggregation step. Given its heaviness in resource consumption, FATE is preferable in powering cross-silo applications instead of cross-device ones.
    \item \textit{FedML}~\cite{he2020fedml}: FedML is also a secure and versatile FL framework that supports distributed mode. Compared to FATE, FedML is more flexible in communication engineering due to the ease of customizing message flow and topology definitions. It is also more lightweight and can thus accommodate training on mobile or IoT devices. Moreover, it can be accelerated with GPUs, while FATE is currently not compatible with hardware accelerators.
    \item \textit{Flower}~\cite{beutel2020flower}: Flower is a concurrent work with FedML, and it concentrates on providing a unified approach for FL with mobile devices. Similar to FedML, Flower bears in mind the goals of being (1) lightweight, (2) extensible, (3) scalable, and (4) compatible with diverse mobile platforms (e.g., Android and iOS) and ML-frameworks (e.g., PyTorch~\cite{paszke2019pytorch} and Tensorflow~\cite{abadi2016tensorflow}). It also implements secure aggregation methods like SecAgg (c.f. \cref{sec:prior_reporting_round}) with easily configurable APIs~\cite{li2021secure}.
    \item \textit{Plato}~\cite{plato}: similar to Flower, Plato also aims to facilitate FL research atop multiple ML backends including PyTorch, Tensorflow, and MindSpore~\cite{mindspore}. Different from its counterparts, Plato is optimized in the development process, e.g., making extensive use of plugin mechanisms to maximize extensibility and maintainability. Apart from lightweight distributed deployment, Plato also offers solutions to integrate secure communication, reverse proxy, and load balancing for best fitting in production environments.
\end{itemize}

Although using production systems yields the most realistic insights, it may not be practical for researchers with limited resources and time budgets. To meet the growing demand for conducting agile FL research, several platforms \textit{that enable system-aware simulation} have been developed. As opposed to system-unaware simulators (e.g., Tensorflow Federated~\cite{tff}, PySyft~\cite{ryffel2018generic}, LEAF~\cite{caldas2018leaf}, OARF~\cite{hu2020oarf}, and FedEval~\cite{chai2020fedeval}), these platforms respect the impact of client system heterogeneity by associating each client with her computation and communication speed, as well as availability dynamics, which are either set manually by developers or by replaying realistic traces. In addition, these platforms also excel in producing comprehensive metrics needed in performance analysis. Compared to real deployment, on the other hand, these systems allow researchers to make fast-forward progress without being blocked by real-world bottlenecks in computation and communication.

\begin{itemize}
    \item \textit{Flower}~\cite{beutel2020flower}: besides deployment on real mobile devices as just mentioned, Flower also supports simulation in the cloud with configurable system-level parameters such as bandwidth constraints and computational capabilities. With that, researchers can experiment with larger and more compute-intensive FL workloads that cannot be run on today's mobile devices.
    \item \textit{FedScale}~\cite{lai2021fedscale}: aside from curating real-world datasets (\cref{sec:cornerstone_benchmarking_datasets}), FedScale also builds an automated runtime to simulate FL in realistic settings. By design, FedScale integrates \texttt{AI Benchmark} and \texttt{MobiPerf Measurements} system traces to simulate clients' heterogeneous training speed and network throughput, respectively. It also incorporates a large-scale user behavior dataset that was formulated in~\cite{yang2021characterizing} to emulate clients' availability dynamics. Compared to Flower, it lacks support for deployment on real distributed devices. Still, it broadly simulates realistic cross-device heterogeneity and can embrace new behavior traces with its APIs.
\end{itemize}

%% file: content/conclusion.tex
%!TEX root=../main.tex
\section{Related Work and Concluding Remarks\label{chap:conclusion}}

\subsection{Related Surveys\label{sec:conclusion_related}}

The motivation for this survey stems from three observations. First, few in-depth surveys focus on system optimization for FL. As FL features strict compliance to privacy regulations as opposed to traditional distributed learning, many survey efforts are directed to the unique challenges such as enforcing data privacy and model robustness~\cite{anirban2019privacy, rigaki2020survey, lyu2020privacy, blanco2020achieving, truong2020privacy}, while the system optimization issues receive less attention in dedicated surveys.

As for common system issues shared by FL and traditional distributed learning, there do exist extensive surveys with detailed discussion. In~\cite{gu2019distributed}, the authors discuss the realm of mobile distributed machine learning, where algorithms are classified into three categories: 1) machine learning optimizers, 2) distributed optimization algorithms, and 3) data aggregation methods. In \cite{tang2020communication}, the authors provide a detailed survey of communication-efficient distributed training algorithms. They identify four key factors that affect the communication cost in distributed learning and organize the literature accordingly: 1) synchronous scheme, 2) system architecture, 3) compression techniques and 4) parallelism of communication and computation.~\cite{shi2020quantitative} also studies the communication issues in distributed deep learning, however, with a focus on the commonly used lossless methods. It also contains a quantitative analysis of these methods based on real-world experiments. Compared to our survey, the scope of these surveys is not narrowed down to FL and thus does not fully capture all the system optimization challenges that are unique in FL. Notably, FL has in its standard workflow the selection phase which needs particular investigation due to client heterogeneity, while traditional distributed learning does not even have the notion of clients.

We notice that there are comprehensive surveys that cover a wide range of FL techniques including system-level optimizations. In~\cite{kairouz2019advances}, the authors discuss in-depth the advances and open problems in FL at the time of writing. The presented problems and solutions are presented in four categories: 1) efficiency and effectiveness, 2) privacy preservation, 3) robustness enforcement and 4) fairness establishment. In~\cite{lim2020federated}, the authors extensively introduce the challenges and research directions of FL at and for mobile edge network. As for FL at mobile edge network, they particularly elaborate techniques on 1) communication efficiency, 2) resource allocation, and 3) privacy and security. In~\cite{wang2021field}, the authors provide recommendations on algorithm-level optimization of FL applications, while they also briefly sketch the system-level constraints and practices. They discuss the reduction of communication, computation, and memory costs, as well as propose a basic model to estimate the communication efficiency of cross-device FL deployment. In~\cite{wahab2021federated}, the authors present a fine-grained multi-level classification of the FL literature, spanning six major topics: 1) statistical challenges, 2) communication efficiency, 3) client selection and scheduling, 4) security concerns, 5) privacy concerns, and 6) service pricing. Compared to this survey, the above-mentioned work is neither intended to focus on system-level optimizations nor do they categorize the literature by the phases in the synchronous training process. The survey most closely related to ours is~\cite{xu2021asynchronous} which is dedicated to summarizing system-level optimizations used in asynchronous FL. As their studied architecture is fundamentally different from ours, their survey could be a good complement to this survey. It is also noteworthy that their classification mechanism on existing techniques is built upon the type of client heterogeneity, which also differs from our taxonomy.

In short, this survey aims to provide a succinct yet complete view of an essential domain in the FL literature: system optimization in synchronous federated training. We also summarize in Tab.~\ref{tab:survey} the main similarities and differences between our survey and the existing surveys that are closely related.

\begin{table*}[h]
    \begin{center}
        \caption{Comparative summary between our survey and the most related work.}
    \label{tab:survey}
    \resizebox{1.0\linewidth}{!}{%
    \begin{tabular}{cm{9cm}m{9cm}}
        \toprule
        Survey & \multicolumn{1}{c}{Similarities} & \multicolumn{1}{c}{Differences} \\
        \midrule
        Gu et al.~\cite{gu2019distributed}
        &
        Similar to our survey, this survey addresses some common challenges of synchronous federated training such as communication efficiency.
        & \hspace{-5mm}
        \begin{minipage}[b]{0.52\textwidth}
            \vspace{1mm}
            \begin{itemize}
            \item As for machine learning optimizers, it does not discuss the server-side ones which are FL-specific.
            \item Our work addresses additional challenges that are not mentioned in this survey such as participant selection and aggregation efficiency.
            \end{itemize}
            \vspace{1mm}
        \end{minipage} \\
        Tang et al.~\cite{tang2020communication}
        & Similar to our survey, this survey addresses some common challenges of synchronous federated training such as communication efficiency.
        & \hspace{-5mm}
        \begin{minipage}[b]{0.52\textwidth}
            \vspace{1mm}
            \begin{itemize}
            \item It considers distributed vanilla SGD, which is different in convergence characteristics from FL, a special case of local SGD.
            \item Our work addresses additional challenges that are not mentioned in this survey such as participant selection and client bias reduction.
            \end{itemize}
            \vspace{1mm}
        \end{minipage} \\
        Shi et al.~\cite{shi2020quantitative}
        & Similar to our survey, this survey addresses some common challenges of synchronous federated training such as communication efficiency.
        & \hspace{-5mm}
        \begin{minipage}[b]{0.52\textwidth}
            \vspace{1mm}
            \begin{itemize}
            \item It focuses on controlling the system architecture and scheduling algorithms in data center learning, which is not applicable to synchronuous FL.
            \item Our work addresses additional challenges that are not mentioned in this survey such as participant selection and data heterogeneity.
            \end{itemize}
            \vspace{1mm}
        \end{minipage} \\
        Kairouz et al.~\cite{kairouz2019advances}
        & Similar to our survey, this survey addresses some common challenges of synchronous federated training such as client bias reduction and communication efficiency.
        & \hspace{-5mm}
        \begin{minipage}[b]{0.52\textwidth}
            \vspace{1mm}
            \begin{itemize}
            \item The scope of this survey is quite board, where for example theoretical analysis of convergence and privacy-preserving mechanisms are also extensively discussed.
            \item Our work addresses additional challenges that are not mentioned in this survey such as participant selection and load balancing.
            \end{itemize}
            \vspace{1mm}
        \end{minipage} \\
        Lim et al.~\cite{lim2020federated}
        & Similar to our survey, this survey addresses some common challenges of FL such as communication efficiency and participant selection. It also mentions the standard workflow of synchronous training.
        & \hspace{-5mm}
        \begin{minipage}[b]{0.52\textwidth}
            \vspace{1mm}
            \begin{itemize}
            \item The scope of this survey is quite board, where many applications of FL in mobile edge network, e.g., cyberattack detection, are also elaborated.
            \item They do not organize system-level optimizations by the phases in the standard training workflow.
            \item Our work addresses additional challenges that are not mentioned in this survey such as client bias reduction and lightweight privacy-preserving aggregation.
            \end{itemize}
            \vspace{1mm}
        \end{minipage} \\
        Wang et al.~\cite{wang2021field}
        & Similar to our survey, this survey addresses some common challenges of FL such as communication efficiency and lightweight privacy-preserving aggregation.
        & \hspace{-5mm}
        \begin{minipage}[b]{0.52\textwidth}
            \vspace{1mm}
            \begin{itemize}
            \item It centers on the algorithm-level optimizations as well as their theoretical guanrantees, while the challenges and directives of system-level issues are briefly introduced.
            \item Our work addresses additional challenges that are not mentioned in this survey such as participant selection and load balancing.
            \end{itemize}
            \vspace{1mm}
        \end{minipage} \\
        Wahab et al.~\cite{wahab2021federated}
        & Similar to our survey, this survey addresses some common challenges of FL such as participant selection and communication efficiency.
        & \hspace{-5mm}
        \begin{minipage}[b]{0.52\textwidth}
            \vspace{1mm}
            \begin{itemize}
            \item The scope of this survey is quite board, where even training service pricing are also included.
            \item Our work contains additional exploratory work in the system community that is not mentioned in this survey including measurement studies and benchmarking suites.
            \end{itemize}
            \vspace{1mm}
        \end{minipage} \\
        Xu et al.~\cite{xu2021asynchronous}
        & Similar to our survey, this survey focuses on optimizing FL training process in the precence of client heterogeneity.
        & \hspace{-5mm}
        \begin{minipage}[b]{0.52\textwidth}
            \vspace{1mm}
            \begin{itemize}
            \item Its target architecture is asynchornous and thus the intersection of its introduced techniques and ours are fairly small.
            \item Our work sorts the literature by the phases in the standard synchronous training workflow, which is distinct from the classification mechanism used in this survey.
            \end{itemize}
            \vspace{1mm}
        \end{minipage} \\
        \bottomrule
    \end{tabular}
    }
    \end{center}
\end{table*}

\subsection{Future Research Directions\label{sec:conclusion_future}}

The primary goal of this survey is to help researchers design future optimization solutions. To stimulate more directives for FL practitioners, we discuss in the following some possible future directions that we derive from the literature as well as our development practice.

\subsubsection{On the Selection Phase\label{sec:conclusion_future_selection}}

% Some RL-based client selection strategies (e.g., Oort~\cite{lai2020oort}) are evaluated on some system-aware simulators which replay the synthetic traces that do not result from real FL training. It thus remains unclear whether their simulation is realistic enough and how their proposed approaches can be improved in emulated environments or industrial practice.

While RL agents are commonly used in guiding client selection for their beneficial balance between exploitation and exploration~\cite{wang2020optimizing, kim2021autofl}, their training costs are inherently high. Before an RL agent is ready for use, it has to learn from tens to hundreds of full-sized FL training processes, whose cost may not be justifiable in practice. Inspired by the experience from container management~\cite{li2021george}, we are interested in whether certain techniques, e.g., transfer learning~\cite{pan2009survey}, can be adopted so that an RL agent can be trained from prior FL tasks, instead of learning from scratch for every single task.

Moreover, existing algorithms all assume the participation scale (i.e., the number of clients training simultaneously) to be the magnitude of a hundred, because involving more clients in a round is observed to have marginal benefits under primary aggregation methods (e.g., FedAvg~\cite{mcmahan2017communication}). On the other hand, the number of available clients at each minute can be as many as thousands in the cross-device practice. It is still desirable to explore the selection of a larger crowd for boosting the time-to-accuracy performance.

\subsubsection{On the Configuration Phase\label{sec:conclusion_future_configuration}}

Aligned with the observations in the literature~\cite{lai2021fedscale}, most prior arts consistently configure different clients. Although there exist some heterogeneity-aware efforts like load balancing where the number of batches, batch size, and number of local epochs can vary across clients (\cref{sec:prior_configuration_latency}), we anticipate that the design space for heterogeneity-aware client configurations could be larger, e.g., using different compression ratios or synchronization frequencies.

To achieve efficient and secure communication, it is also worth studying how to combine sparsification techniques with privacy-preserving methods. For example, model sparsification is by nature not compatible with secure multi-party computation protocols such as SecAgg~\cite{bonawitz2017practical}, because the coordinates of sparsified model values often vary across clients, preventing the pairwise masks from being canceled out during model aggregation as required by SecAgg. A recent work named SparseSecAgg~\cite{ergun2021sparsified} attempts to tackle this issue, however, it implies the need for sharing some sparsified locations between each pair of clients, which cannot be directly extended to conventional sparsification schemes such as the top $s\%$ scheme (\cref{sec:prior_configuration_size}). More general reconciliation between the two techniques is still ripe for future investigation.

\subsubsection{On the Reporting Phase\label{sec:conclusion_future_reporting}}

As the system bottleneck is usually assumed to locate in clients instead of the server, most of the existing optimization efforts focus on improving the utility (system and statistical) of clients. It is thus interesting to investigate whether such an assumption holds in practice, especially when the scalability of the server is restricted due to rigid capabilities or limited budgets, or when its aggregation latency is not negligible due to the presence of large-scale models. 

Besides, existing lightweight privacy-preserving aggregation methods are not compatible with robustness enforcement techniques~\cite{kairouz2019advances}. Because the true values of local models are concealed from the server, it has no way of inspecting their numerical features~\cite{blanchard2017machine, chen2017distributed, guerraoui2018hidden, yin2018byzantine} or validating their performance for anomaly detection~\cite{liu2018fine, wang2019neural, fang2020local}. This raises the tension between privacy and robustness. Leveraging Trusted Execution Environment (TEE) \cite{sabt2015trusted, jauernig2020trusted} to perform model inspection securely might be a helpful workaround. However, due to current limits on TEE's hardware capabilities \cite{orenbach2019cosmix}, there could be a foreseeably large performance downgrade of doing so. It thus remains largely unexplored as to how to navigate the sweet point of jointly maximizing accuracy, performance, privacy, and robustness in synchronous federated training.

\subsection{Discussion\label{sec:conclusion_dicussion}}

We further discuss the applicability of the above-introduced optimization techniques in different contexts. 

\subsubsection{Cross-Device FL and Cross-Silo FL\label{sec:conclusion_dicussion_cross}}

FL applications are often categorized as either cross-device FL (where the participants are a mass of less capable mobile or IoT devices) or cross-silo FL (where the participants are 2-100 organizational entities)~\cite{kairouz2019advances}. While the FL workflow that we base on throughout this survey is primarily proposed for cross-device FL~\cite{bonawitz2019towards, yang2021characterizing, lai2021fedscale}, it also generalizes to cross-silo settings. Hence, the scope of this survey does not preclude cross-silo FL, and hence many practical methods mentioned here should apply to both settings. For those techniques that are suitable for merely one setting, we have emphasized their limitations and stated the practical reasons behind them.

\subsubsection{Horizontal FL and Vertical FL\label{sec:conclusion_dicussion_horizontal}}

Based on the characteristics of data distribution, FL applications can also basically be classified as horizontal FL (where clients' data have the same feature space but different samples) or vertical FL (where clients' data share the same sample space but have different features)~\cite{yang2019federated}. Horizontal FL is typically initiated by a service provider who wants to improve the quality of an ML application to better fit the end users' data of a specific domain. For example, by combining the statistics of users' input habits, the developers of a mobile keyboard application could achieve a more intelligent prediction of the next words typed by the users, thus enhancing their experience.

On the other hand, vertical FL participants are usually a few organizations who hold data in different domains while being likely to achieve win-win cooperation by knowledge sharing. For instance, a regional e-commerce company and a bank may share the same set of clients. By incorporating the knowledge of the clients' revenue and expenditure recorded at the bank as well as the purchasing traces collected by the e-commerce, they can build a more accurate prediction model on the clients' purchasing behaviors, benefiting both of their businesses.

Given that there exist diverse training workflows of vertical FL and the community has not yet achieved a consensus on the use of any one~\cite{yang2019quasi, hu2019fdml, gu2020federated, cheng2021secureboost}, much discussion in this survey is biased towards horizontal FL that has a widely acknowledged architecture (\cref{sec:background_federated}). Still, we can provide some insights for optimizing vertical FL training in general:

\begin{enumerate}
    \item As the data is feature-partitioned, vertical FL needs the participation of all clients in each model update attempt. Thus, optimizations on participant selection (\cref{sec:prior_selection}) are not applicable to vertical FL.
    \item In terms of communication, model-agnostic compression techniques (\cref{sec:prior_configuration_size}) should still apply to plaintext variants of vertical FL such as~\cite{feng2020multi}. However, synchronization frequency reduction (\cref{sec:prior_configuration_freq}) is not possible in vertical FL, as each client does not have a complete model and cannot conduct training independently.
    \item As for local training, as all participants need processing data of the same sample space, load balancing techniques that assign a different amount of data samples to each client (\cref{sec:prior_configuration_latency}) are not feasible in vertical FL. Moreover, as clients inherently have different local models, adaptive optimization and bias reduction methods (\cref{sec:prior_configuration_round}) from horizontal FL do not generalize well to vertical FL. 
    \item Unlike horizontal FL, the forms of aggregation in vertical FL differ significantly by the model architecture. Thus, the mentioned optimizations (\cref{sec:prior_reporting}) can hardly be a generic remedy.
\end{enumerate}

Besides horizontal FL and vertical FL, there recently emerges a novel collaborative learning paradigm called Federated Transfer Learning (FTL)~\cite{yang2019federated, liu2020secure} which can cope with more relaxed assumptions of client data distribution using transfer learning techniques~\cite{pan2009survey}. Specifically, it deals with the cases where two clients' data not only differ in sample space but also feature space. Toward this end, FTL learns a common feature representation between the two clients and minimizes the empirical errors in predicting one client's labels by leveraging the labels of other clients. As FTL's training workflow ensembles some variants of vertical FL~\cite{yang2019federated}, we believe that the above-mentioned insights for vertical FL also hold for FTL.

\subsection{Conclusion\label{sec:conclusion_conclusion}}

In this survey, we focus on system optimization in synchronous federated training and propose a natural taxonomy that categorizes existing solutions based on both the training phase and the type of utility at which they target. Apart from problem-driven attempts, we also include related cornerstone efforts including measurement studies and benchmarking suites. We expect this manuscript to be a \textit{useful guideline} for the design and implementation of federated learning systems.